\documentclass[epj]{svjour}
\usepackage{graphics}
\usepackage{psfig}
\def\be{\begin{equation}}
\def\ee{\end{equation}}
\def\bea{\begin{eqnarray}}
\def\eea{\end{eqnarray}}
\newcommand{\beq}{\begin{equation}}
\newcommand{\eeq}[1]{\label{#1} \end{equation}}

\newcommand{\insertplot}[1]{\centerline{\psfig{figure={#1},width=8.4cm}}}

\begin{document}
\title{Freeze out of the expanding system}
\author{V.K. Magas \inst{1}
\and L.P. Csernai \inst{2,3}
\and E. Moln\'ar \inst{2}
}                     
\institute{Departament d'Estructura i Constituents de la Mat\'eria,
 Universitat de Barcelona, Diagonal 647, 08028 Barcelona, Spain.
\and Section for Theoretical and Computational Physics,
     University of Bergen, Allegaten 55, 5007 Bergen, Norway.
     \and MTA-KFKI, Research Inst of Particle and Nuclear Physics, H-1525 Budapest 114, P.O.Box 49, Hungary.
}
\date{Oct. 17, 2006}
%
\abstract{
The freeze out (FO) of the expanding systems, created in relativistic heavy ion collisions, is discussed. We start with kinetic FO model, which realizes complete physical FO in a layer of given thickness, and then combine our gradual FO equations with Bjorken type system expansion into a unified model. We shall see that the basic FO features, pointed out in the earlier works, are not smeared out by the expansion. 
} 
\maketitle

At highest energies available nowadays in relativistic heavy ion collisions at RHIC the total number of the produced particles exceeds 6000, 
therefore one can expect that the produced system behaves as a "matter" and generates collective effects. Indeed strong collective flow patterns have been measured at RHIC, 
which suggests that  
the hydrodynamical models are well justified during the intermediate stages of the reaction: from the time when local equilibrium is reached until the freeze out (FO),
when the hydrodynamical description breaks down.
During this FO stage, the matter becomes so dilute and cold that  particles stop interacting and
stream towards the detectors freely, their momentum distribution freezes out.
The FO stage is essentially the last part of a collision process and the main source for observables.

Nowadays, FO is usually simulated in two extreme ways: A) FO on a hypersurface with zero thickness, B) FO described by volume emission model or hadron cascade, which in principle requires an infinite time and space for a complete FO. At first glance it seems that one can avoid troubles with FO modeling using hydro+cascade two module model \cite{hydro_cascade}, since in hadron cascades gradual FO is realized automatically. However, in a such a scenario there is an uncertain point, actually uncertain hypersurface, where one switchs from hydrodynamical to kinetic modeling. First of all it is not clear how to determine such a hypersurface. This hypersurface in general may have both time-like and space-like parts. Mathematically this problem is very similar to hydro to FO phase transition on the infinitely narrow FO hypersurface, therefore for example all the problems discussed for FO on the hypersurface with space-like normal vectors will take place here. Another complication is that while for the post FO domain we have mixture of non-interacting ideal gases, now for the hadron cascade we should generate distributions for the interacting hadronic gas of all possible species, as a starting point for the further cascade evolution. The volume emission models are based on the kinetic equations \cite{vol_em,old_SL_FO_2} 
defining the evolution of the distribution functions, and therefore these also require to generate initial distribution functions for the interacting hadronic species on some hypersurface. 

In this work we present FO model which allows us to study FO in a layer of any thichness, $L$, from $0$ to $\infty$, and which connects the pre FO hydrodynamical quantities, like energy density, $e$, baryon density, $n$, with post FO distribution function in a relatively simple way. Many building blocks of the model are Lorentz invariant and can be applied to both time-like and space-like FO layers. In this work we are going to include Bjorken like expansion in our FO model, in contrast to the older versions \cite{old_SL_FO_2,old_SL_FO,old_TL_FO,Mo05a,Mo05b}. In this latter case the FO layer is a domain restricted by two hypersurfaces $\tau=\tau_1$ and $\tau=\tau_1+L$ ($\tau$ is the proper time).

We are going to review briefly all the steps done to derive our model. We will skip all the detailed derivations, refering to the corresponding publication, and will show and discuss only a small part of results, due to limited space.

Starting from the Boltzmann Transport Equation, introducing two components of the distribution function, $f$: the interacting, $f^i$, and the frozen out,
$f^f$  ones, ($f=f^i+f^f$), and assuming that FO is a directed process (i.e. neglecting the gradients of the distribution functions in the directions perpendicular to the FO direction comparing to that in the FO direction) we can obtain the following system of the equations \cite{Mo05a,ModifiedBTE}: 
\be
\frac{d f^{i}}{d s}  = - \frac{P_{esc} }{\tau_{FO}}f^{i} + \frac{ f_{eq}(s) - f^{i} }{\tau_{th}} \, ,
\quad
\frac{d f^{f}}{d s}  =  \frac{P_{esc}}{\tau_{FO}} f^{i} \, .
\label{sys}
\ee
The FO direction is defined by the unit vector $d\sigma_\mu$. FO happens in a layer of given thickness $L$ with two parallel boundary hypersurfaces perpendicular to $d\sigma_\mu$, and $s=d\sigma_\mu x^\mu$ is a variable in the FO direction. We work in the reference frame of the front, where $d\sigma_\mu$ is either $(1,0,0,0)$ for the time-like FO, or $(0,1,0,0)$ for the space-like FO. The $\tau_{FO}$ is some characteristic
length scale, like mean free path or mean collision time for time-like FO. The rethermalization of the interacting component is taken into account via the relaxation time approximation,
where $f_i$ approaches the equilibrated J\"uttner distribution, $f_{eq}(s)$, with a
relaxation length, $\tau_{th}$. The system (\ref{sys}) can be solved semi-analytically
in the fast rethermalization limit \cite{Mo05a}.

According to the above references, the basis of the model, i.e. the
invariant escape probability within the FO layer of the thickness $L$, 
for both time-like and space-like normal vectors is given as \cite{Mo05a,Mo05b,kemer}
\be
   P_{esc} =
   \left( \frac{L}{L-x^\mu d\sigma_\mu} \right)
   \left(\frac {p^\mu d\sigma_\mu}{p^\mu u_\mu}\right)\
   \Theta(p^\mu d\sigma_\mu)\,,
\label{esc1}
\ee
where $p^\mu$ is a particle four-momentum, $u^\mu$ is the flow velocity.
In fact the model based on the escape rate (\ref{esc1}) is a generalization of a simple kinetic models studied in Refs. \cite{old_SL_FO_2,old_SL_FO,old_TL_FO}, which
can be restored in the $L\rightarrow \infty$ limit.
Here we will concentrate on the time-like case only, where the above $\Theta$ function
is unity. 

Simple semianalytically solvable FO models studied in \cite{old_SL_FO_2,old_SL_FO,old_TL_FO,Mo05a,Mo05b} are missing an important ingredient - the expansion of the freezing out system. The open question is whether the features of the FO, found in those papers, will survive if the system expasion is included. In this work we present a model which includes both gradual FO and Bjorken-like expansion of the system. And we will see that the answer is "yes" - the basic features of the post FO distrubitions will not be smeared out by the expansion.

First, let us remind the reader the basics of the famous Bjorken model. Bjorken model is one-dimensional in the same sense as discussed before eq. (\ref{sys}) - only the proper time, $\tau=\sqrt{t^2-x^2}$, gradients are considered. Here the reference frame of the front, $d\sigma^\mu=(1,0,0,0)$, is the same as the local rest frame, $u^\mu=(1,0,0,0)$. The evolution of the energy density and baryon density is given by the following equations:
\beq
\frac{d e}{d \tau}=-\frac{e+P}{\tau}\,, \quad \frac{d n}{d \tau}=-\frac{n}{\tau}\,,
\eeq{Bjorken}   
where $P$ is the pressure. 
The initial conditions are given at some $\tau=\tau_0$. 

Applying our FO model to such a system, we obtain:
\beq
df^i(\tau')=-\frac{d\tau'}{\tau_{FO}}\frac{L}{L-\tau'}f^i(\tau')+\frac{d\tau'}{\tau_{th}}
\left[f_{eq}(\tau')-f^i(\tau')\right]\,,
\eeq{dfi}
\beq
df^f(\tau')=+\frac{d\tau'}{\tau_{FO}}\frac{L}{L-\tau'}f^i(\tau')\,,
\eeq{dff}
where FO begins at $\tau=\tau_1$ and $\tau'=\tau-\tau_1$. Taking the fast rethermalization limit, similarly to what is done in \cite{old_TL_FO}, we can obtain simplified equations for $f^i$, which is a thermal distribution $f^i(\tau)=f_{eq}(\tau)$, $f^f$ as well as for $e^i, n^i$ and $e^f, n^f$:
\beq
\frac{d e^i}{d \tau'}=-\frac{e^i}{\tau_{FO}}\frac{L}{L-\tau'}\,, \quad 
\frac{d n^i}{d \tau'}=-\frac{n^i}{\tau_{FO}}\frac{L}{L-\tau'}\,,
\eeq{2int}   
\beq
\frac{d e^f}{d \tau'}=+\frac{e^i}{\tau_{FO}}\frac{L}{L-\tau'}\,, \quad 
\frac{d n^f}{d \tau'}=+\frac{n^i}{\tau_{FO}}\frac{L}{L-\tau'}\,.
\eeq{2free}   

Now the idea is to create a system of equations which would describe a fireball which simultaneously expands and freezes out. Let us put our two components ($e=e^i+e^f$) into the first equation of (\ref{Bjorken}) and do some simple algebra:
\begin{equation}
\frac{d e^i}{d\tau}+
\frac{d e^f}{d\tau}=
-\frac{e^i+P^i}{\tau}-\frac{e^f}{\tau}-
\frac{e^i}{\tau_{FO}}\frac{L}{L-\tau'}+\frac{e^i}{\tau_{FO}}\frac{L}{L-\tau'}\,,
\label{fif}
\end{equation}
where last two terms add up to zero; the free component, of course, has no pressure. So far our eq. (\ref{fif}) is completely identical to the first equation of (\ref{Bjorken}). Our assumption is that our system evolves in such a way that eq. (\ref{fif}) is satisfied as a system of two separate equations for interacting and free components \cite{Bjorken_FO}:
\beq
\frac{d e^i}{d \tau}=-\frac{e^i+P^i}{\tau}-\frac{e^i}{\tau_{FO}}\frac{L}{L+\tau_1-\tau}\,, 
\eeq{eint}   
\beq
\frac{d e^f}{d \tau}=-\frac{e^f}{\tau}+\frac{e^i}{\tau_{FO}}\frac{L}{L+\tau_1-\tau}\,.
\eeq{efree}   
Similarly we can obtain equations for baryon density \cite{Bjorken_FO}:
\beq
\frac{d n^i}{d \tau}=-\frac{n^i}{\tau}-\frac{n^i}{\tau_{FO}}\frac{L}{L+\tau_1-\tau}\,, 
\eeq{nint}   
\beq
\frac{d n^f}{d \tau}=-\frac{n^f}{\tau}+\frac{n^i}{\tau_{FO}}\frac{L}{L+\tau_1-\tau}\,.
\eeq{nfree}
   
Thus, finally, we have the following simple model of fireball created in relativistic heavy ion collision.\\
{\bf Initial state, $\tau=\tau_0$:}\ \  $e(\tau_0)=e_0$, $n(\tau_0)=n_0$.\\
{\bf Phase I, Pure Bjorken hydrodynamics, $\tau_0\le \tau\le \tau_1$}
\beq
e(\tau)=e_0\left(\frac{\tau_0}{\tau}\right)^{1+c_o^2}\,, \quad
n(\tau)=n_0\left(\frac{\tau_0}{\tau}\right)\,,
\eeq{pure_bjor}
where $P=c_o^2 e$ - equation of state (EoS) in general form.\\
{\bf Phase II, Bjorken expansion and gradual FO, \linebreak $\tau_1\le \tau\le \tau_1+L$}
\beq
e^i(\tau)=e_0\left(\frac{\tau_0}{\tau}\right)^{1+c_o^2}\left(\frac{L+\tau_1-\tau}{L}\right)^{L/\tau_{FO}}\,, 
\eeq{bjor_FO_1}
\beq
n^i(\tau)=n_0\left(\frac{\tau_0}{\tau}\right)\left(\frac{L+\tau_1-\tau}{L}\right)^{L/\tau_{FO}}\,.
\eeq{bjor_FO_2}
With these last equations we have completely determined evolution of the interacting component \cite{Bjorken_FO}. Knowing $e^i(\tau)$ and EoS we can find temperature, $T_i(\tau)$. Due to symmetry of the system $u_i^\mu(\tau)=u^\mu(\tau_0)=(1,0,0,0)$. Finally, $f^i(\tau)$ is a thermal distribution with given $T_i(\tau)$, $n^i(\tau)$, $u_i^\mu(\tau)$. 

However for us the more interesting is free component, which is the source of the observables. Eqs. (\ref{eint},\ref{efree}) give us the evolution of the $e_f$ and $n_f$, and one can easily check that these two equations are equivalent to the following equation on the distribution function:
\beq
\frac{d f^f}{d \tau}=-\frac{f^f}{\tau}+\frac{f^i}{\tau_{FO}}\frac{L}{L+\tau_1-\tau}\,.
\eeq{ffree}
The measured post FO spectra are given by $f^f(L+\tau_1)$.

Aiming for a qualitative illustration of the FO process we show below the results for the massless ideal gas without conserved charges with J\"uttner equilibrated distribution ($P^i=e^i/3$, $e^i=\frac{3}{\pi^2} T_i^4$, $f^i(\tau,|\vec{p}|)=\frac{1}{(2\pi)^3}e^{-|\vec{p}|/T_i(\tau)}$). We have taken the following values of the parameters: $\tau_0=0.05\ fm$, $T_i(\tau_0)=835\ MeV$; $\tau_1=5\ fm$, $T_i(\tau_1)=T_{FO}=180\ MeV$; $\tau_{FO}=0.5\ fm$ and we present results for different values of FO time $L$. 

\begin{figure}[htb!]
                \insertplot{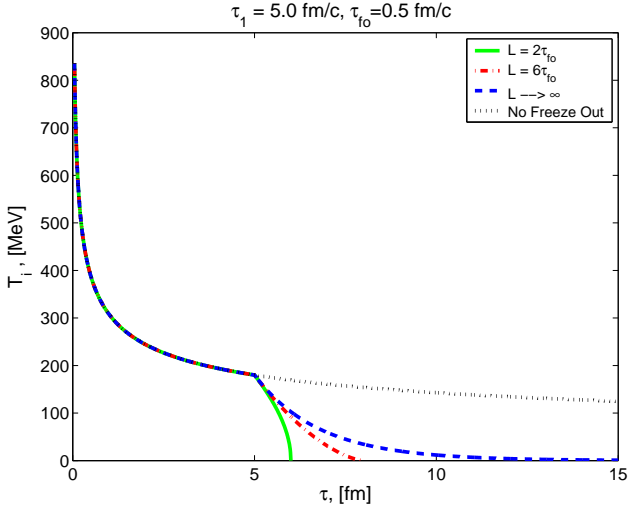}
\caption{Evolution of the temperature of the interacting matter for
different FO layers. $T_i(\tau_0=0.05\ fm)=835\ MeV$, $T_{FO}=180\ MeV$. "No Freeze Out" means that we used standart Bjorken hydrodynamics even in phase II.}
\label{fig1}
\end{figure}

\begin{figure}[htb]
                 \insertplot{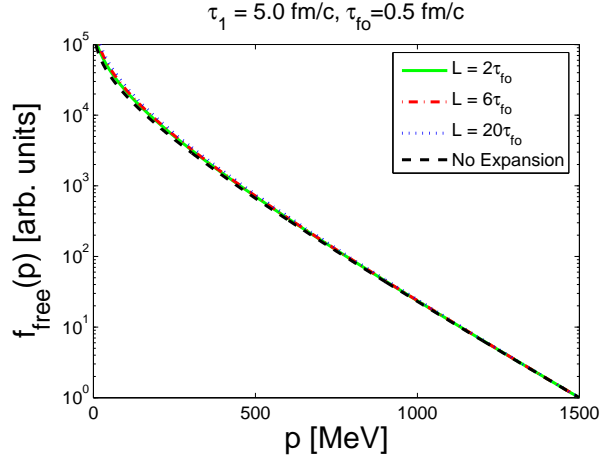}
\caption{Final post FO distribution for different FO layers as a function of the momentum in the
FO direction, $p=p^x$ in our case ($p^y=p^z=0$). The initial conditions are specified in the text. "No Expansion" curve
is given by the analytical expression \cite{old_TL_FO}: $f^f(p)=-\frac{4}{(2\pi)^3}Ei(-\frac{p}{T_{FO}})$.}
\label{fig2}
\end{figure}

Fig. \ref{fig1} shows the evolution of the temperature of the interacting matter.  
As it was already shown in \cite{old_TL_FO,Mo05b} the final post FO particle distributions
are non-equilibrated distributions, which deviate from thermal ones
particularly in the low momentum region.
By introducing  and varying the thickness of the FO layer, $L$, we are strongly affecting the evolution
of the interacting component, see Fig. \ref{fig1}, but we gain see the universality of the final post FO distribution: 
for $L>2\tau_{FO}$ it already looks very close to that for an infinitely long FO calculations: see Fig. \ref{fig2}.
The inclusion of the expansion into our consideration does not smear out this very important feature of FO.
More results can be found in \cite{Bjorken_FO}.

In our opinion these results may justify the use of FO hypersurface in hydrodynamical models for heavy ion collisions,
but with a proper non-thermal post FO distributions.
If the FO layer is thick enough, say $L>2\tau_{FO}$, then it doesn't matter
how thick was FO layer, we do not need to model the FO dynamics in details.
Once we have a good parameterization of the post FO spectrum (still asymmetric, non-thermal), for example analytical post FO distribution obtained in Ref. \cite{old_TL_FO} (see caption of Fig. \ref{fig2}), then the parameters of this distribution can be found from the conservation laws, as it is usually done for sharp FO, with some volume scaling factor to effectively account for the expansion during FO.  It is important to always check the non-decreasing entropy
condition \cite{Bjorken_FO,cikk_2} to see whether such a process is physically possible.


\end{document}